%% file: main.tex
\documentclass[screen, nonacm]{acmart}
\usepackage{array}
\usepackage{booktabs}
\usepackage{multirow}
\usepackage{longtable}
\AtBeginDocument{%
  }

\acmISBN{978-1-4503-XXXX-X/2018/06}

\begin{document}

\title{Beyond Overt Reactions: Analyzing Subtle User Emotional Response to Unexpected In-Vehicle System Behavior}


\author{Huy Quyen Ngo}
\authornote{Both authors contributed equally to this research.}
\email{huyquyen@andrew.cmu.edu}
\author{Suresh Kumaar Jayaraman}
\authornotemark[1]
\email{sureshkj@andrew.cmu.edu}
\affiliation{%
  \institution{Carnegie Mellon University}
  \city{Pittsburgh}
  \state{Pennsylvania}
  \country{USA}
}

\author{Brian Mok}
\affiliation{%
  \institution{BMW Group Technology Office USA}
  \city{Mountain View}
  \state{California}
  \country{USA}}
\email{Brian.Mok@bmwgroup.com}

\author{Ken Friedl}
\affiliation{%
  \institution{BMW Group Germany}
  \city{Munich}
  \state{Bavaria}
  \country{Germany}}
\email{ken.friedl@bmw.de}

\author{Oliver Krause}
\affiliation{%
  \institution{BMW Group Germany}
  \city{Munich}
  \state{Bavaria}
  \country{Germany}}
\email{Oliver.A.Krause@bmw.de}

\author{Aaron Steinfeld}
\affiliation{%
  \institution{Carnegie Mellon University}
  \city{Pittsburgh}
  \state{Pennsylvania}
  \country{USA}}
\email{as7s@andrew.cmu.edu}

\author{Nikolas Martelaro}
\affiliation{%
  \institution{Carnegie Mellon University}
  \city{Pittsburgh}
  \state{Pennsylvania}
  \country{USA}}
\email{nmartela@andrew.cmu.edu}




\renewcommand{\shortauthors}{Ngo and Jayaraman et al.}

\begin{abstract}
Modern vehicles, with advanced AI voice and autonomous navigation features, extend beyond traditional driving but, like any autonomous system, can potentially make mistakes or behave in ways unexpected by users. Although providing real-time explanations can alleviate some confusion, constant information can overwhelm users and potentially cause unnecessary distractions. Some situations may require explanations or corrective vehicle behavior, and thus, recognizing user response to unexpected vehicle behavior is critical. To investigate such user responses, our study focused on collecting and analyzing user behavioral responses to unexpected events while interacting with a fully autonomous vehicle in a driving simulator. We also aimed to address the lack of datasets capturing subtle user responses (facial, spoken language, physiological signals) to in-vehicle events, as existing datasets primarily focus on strong emotional signals in conventional human-driven cars and user response to external road and traffic conditions. Users were exposed to stimuli designed to induce surprise, confusion, and frustration while performing a secondary task on a tablet and interacting with the vehicle through voice commands and in-vehicle displays. We collected a multi-modal dataset with video, audio, and heart rate data and gained insights into subtle user responses that underscored the need for further investigation of nuanced user behaviors. These observations highlight the importance of designing vehicles that recognize and adapt to occupants’ behavior, potentially improving their experience.
\end{abstract}

\begin{teaserfigure}
    \centering
\includegraphics[width=0.95\textwidth, trim=0 8.5cm 0 0, clip]{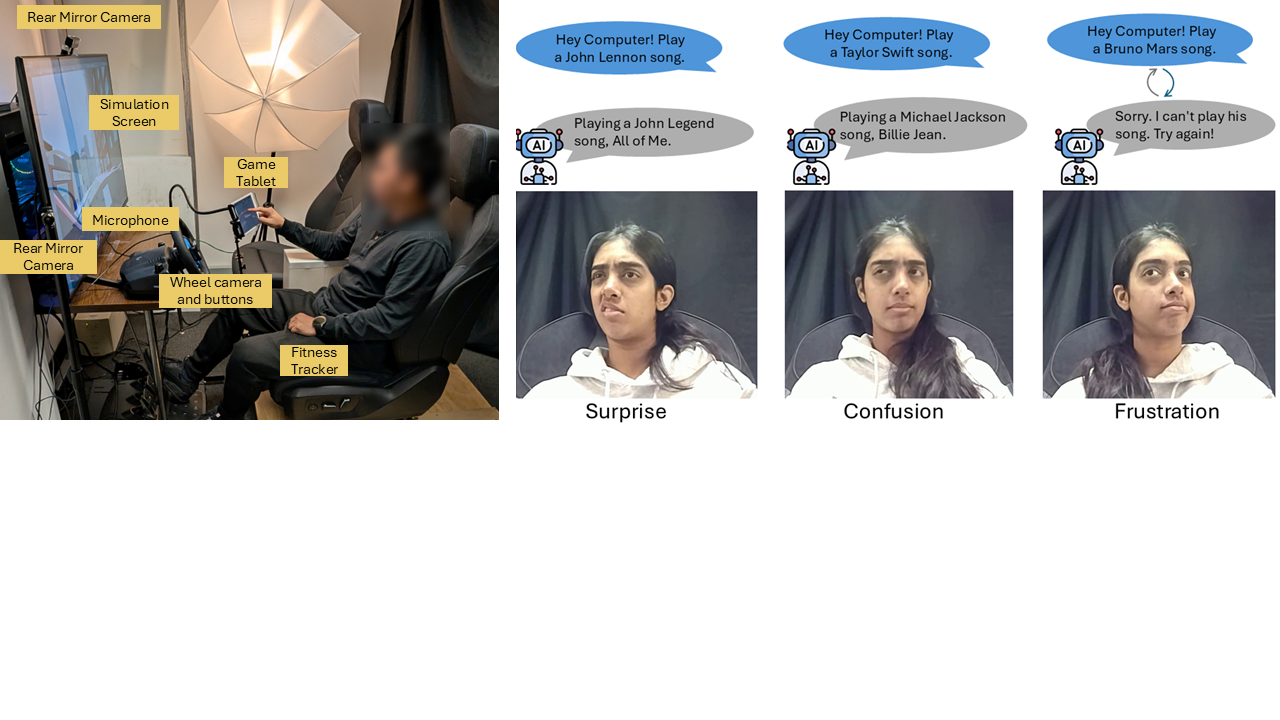}
  \caption{The driving simulator setup with a word puzzle game on a tablet and devices to capture user response. (Left). A user shows distinct facial expressions in response to different unexpected system behaviors (Right).}
  \Description{}
  \label{fig:teaser}
\end{teaserfigure}



\maketitle

\input{text/1_intro}

\input{text/2_background}

\input{text/4_study}

\input{text/5_results}

\input{text/7_discussion}

\input{text/8_conclusion}




\bibliographystyle{ACM-Reference-Format}
\bibliography{main_bib}

\clearpage
\appendix

\section{Appendix}

\subsection{Stimuli in Scenarios} \label{stimuli-in-scenarios}

\begin{figure}[htp]
    \centering
    \includegraphics[width=0.325\linewidth]{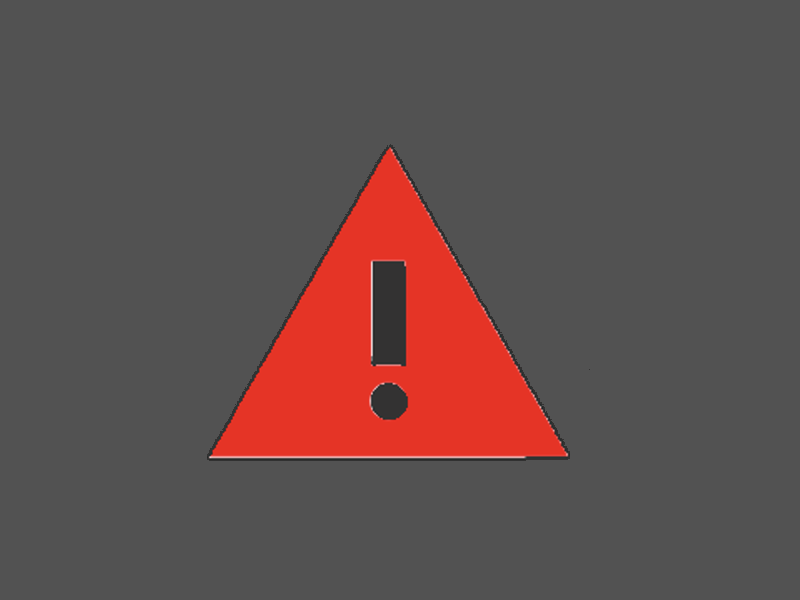}
    \includegraphics[width=0.325\linewidth]{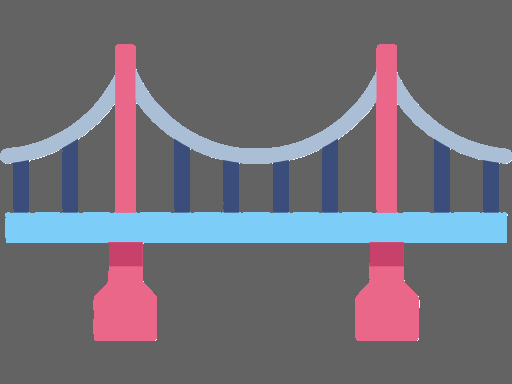}
    \includegraphics[width=0.31\linewidth]{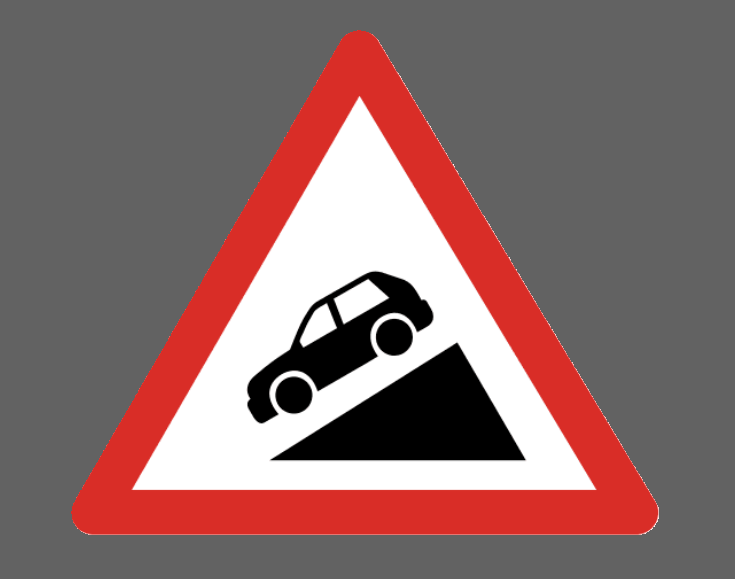}
    \caption{Alert icons for the surprise (left), confusion (middle), and frustration (right) response triggering stimuli.}
    \label{fig:alert-icon}
\end{figure}

\begin{figure}[htp]
    \centering
    \includegraphics[width=0.35\linewidth]{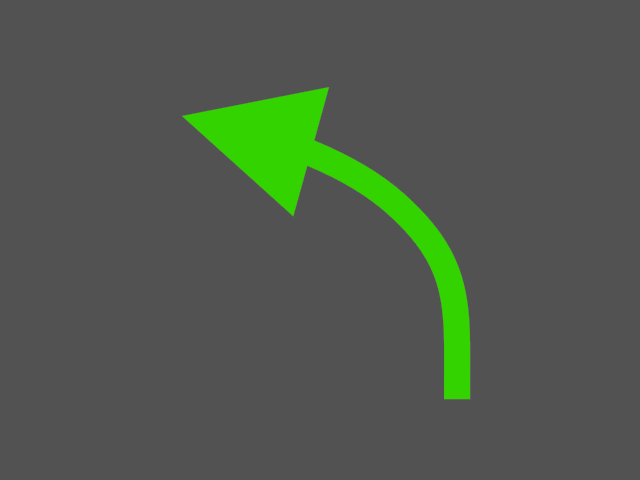}
    \includegraphics[width=0.35\linewidth]{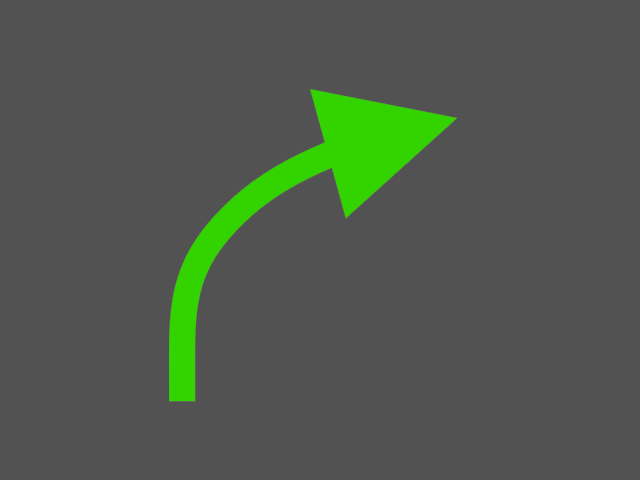}
    \includegraphics[width=0.35\linewidth]{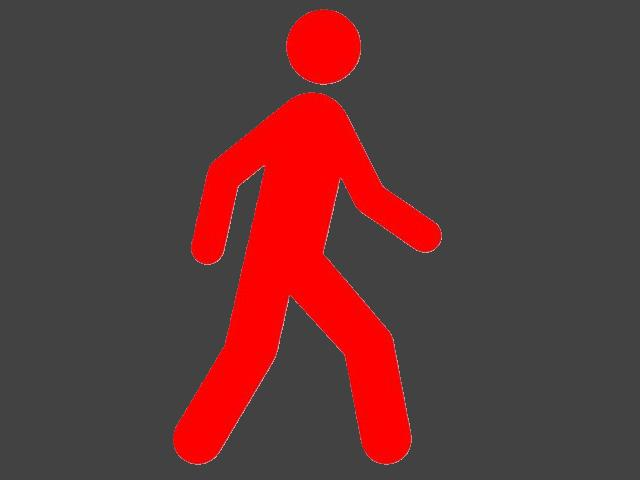}
    \includegraphics[width=0.35\linewidth]{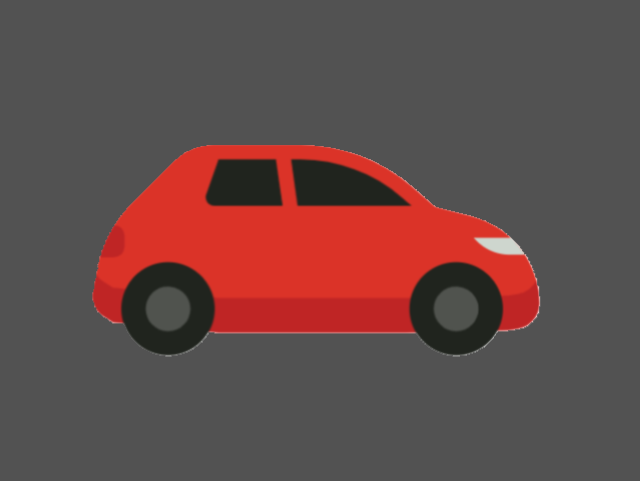}
    \caption{Alert icons for the correct stimuli, namely left turn (upper left), right turn (upper right), pedestrian ahead alert (bottom left), and car ahead alert (bottom right).}
    \label{fig:alert-icon-correct}
\end{figure}

The stimuli in our three scenarios are described in Table \ref{tab:stimuli-description}. The alert icons for the unexpected stimuli and the correct stimuli are shown in Figures \ref{fig:alert-icon} and \ref{fig:alert-icon-correct}, respectively.
For the alert scenario, participants had to press a button on the wheel after they had understood the alert for the 3-question survey to appear on the tablet.

\begin{longtable}{|p{2cm}|p{1.5cm}|p{11cm}|}
\hline
\textbf{Scenario} & \textbf{Response} & \textbf{Stimuli Description}\\
\hline
\multirow{3}{*}{} & Surprise & Participant requested songs from John Lennon. System played a John Legend song. \\
\cline{2-3}
& Confusion & Participant requested songs from Taylor Swift. System played a Michael Jackson song. \\
\cline{2-3}
Music Playing & Frustration & Participant requested songs from Bruno Mars. System said ``Sorry. I can't play his songs'' and proceeded to turn on the weather channel. Stimulus was repeated for maximum 3 times participants request, or 45 seconds after the first request, whichever came first. At the fourth time or after 45 seconds, the system said ``Sorry. I can't play his songs. Please try again in a while''. \\
\hline
\multirow{3}{*}{} & Surprise & Participant ordered sushi. System ordered sushi, but with extra \$19.55 delivery fee. \\
\cline{2-3}
& Confusion & Participants ordered pizza. System ordered fried chicken. \\
\cline{2-3}
Food Ordering & Frustration & Participant ordered taco. System says ``Could not place the order. Can you try again?''. The stimulus was repeated for maximum 3 times participants order, or 45 seconds after the first order, whichever came first. At the fourth time or after 45 seconds, the system said ``Could not place the order. Please try again in a while''. \\
\hline
\multirow{3}{*}{} & Surprise & A warning icon was shown, but nothing is on the screen. Only after a few seconds, some trash cans were seen in the middle of the road, and the vehicle steered to avoid them. The icon is shown in Figure \ref{fig:alert-icon} (left). \\
\cline{2-3}
Alert & Confusion & A bridge icon was shown, but no bridge was found anywhere. The icon is shown in Figure \ref{fig:alert-icon} (middle). \\
\cline{2-3}
& Frustration & A steep slope icon was shown, but no steep slope was found anywhere. At the same time, the system repeatedly said ``Steep slope ahead. Pay attention'' and could not be turned off. The icon is shown in Figure \ref{fig:alert-icon} (right). \\
\hline
\caption{Stimuli Description}
\label{tab:stimuli-description}
\vspace{-2mm}
\end{longtable}

\subsection{Tablet game}

\begin{figure}[htp]
    \centering
    \includegraphics[width=0.8\linewidth]{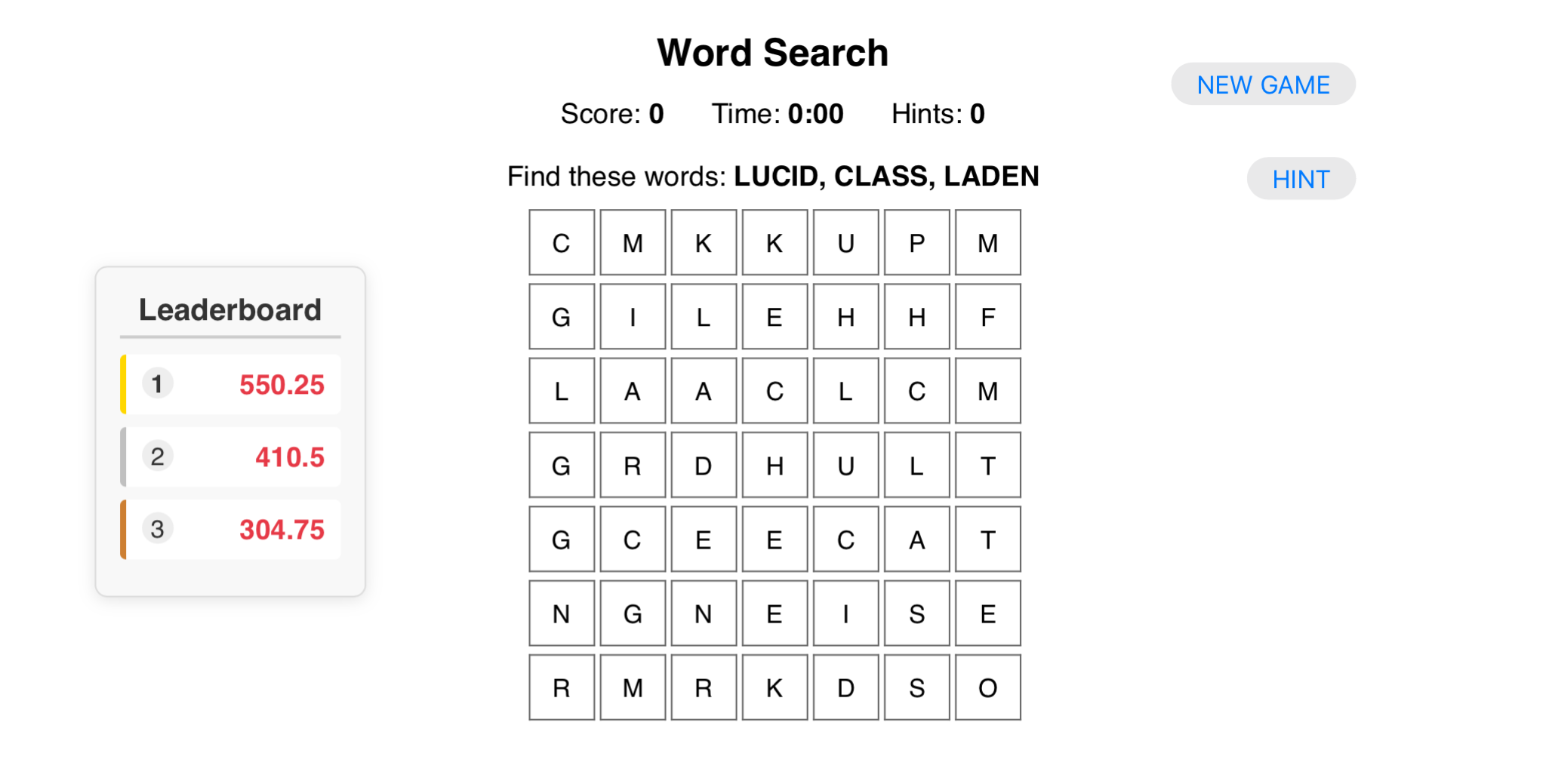}
    \caption{The screen of the tablet game.}
    \label{fig:tablet-game}
\end{figure}

The word search game on the tablet is shown in Figure \ref{fig:tablet-game}.

Participants needed to find all the words in the grid. The words changed in every game, and the level of difficulty varied. Participants could use hints or start a new game. The leaderboard showed high scores of previous participants.

\subsection{Other Surveys in the study} \label{survey}
\subsubsection{Pre-Study Survey (Scale is from “Strongly Disagree” to “Strongly Agree”)} \

\begin{enumerate}
    \item I am familiar with technologies in general (smart cars, robots, smart homes, etc).
    \item I am familiar with or have experience playing video games.
    \item I am familiar with autonomous vehicles.
    \item I am familiar with voice-based agents (Amazon Alexa, Google Homes, Apple Siri, etc).
\end{enumerate}

\subsubsection{NASA-TLX Survey (Scale is from “Extremely Low” to “Extremely High”)} \

\begin{enumerate}
    \item How mentally demanding was the task?
    \item How physically demanding was the task?
    \item How hurried or rushed was the pace of the task?
    \item How successful were you in accomplishing what you were asked to do?
    \item How hard did you have to work to accomplish your level of performance?
    \item How insecure, discouraged, irritated, stressed, and annoyed were you?
\end{enumerate}

\subsubsection{Demographics} \

\begin{enumerate}
    \item What is your age? 
    \item What is my gender? 
        \begin{enumerate}
            \item Male
            \item Female
            \item Other
            \item I prefer not to answer the question
        \end{enumerate} 
    \item Race / Ethnicities?
        \begin{enumerate}
            \item American Indian / Alaska Native
            \item Asian
            \item African American
            \item Native Hawaiian / Other Pacific Islander
            \item White
            \item Other
            \item Unknown
            \item I prefer not to answer the question
        \end{enumerate} 
    \item What is your highest education (pursuing or completed)?
        \begin{enumerate}
            \item Less than high school degree
            \item High school graduate (high school diploma or equivalent including GED)
            \item Some college but no degree
            \item Associate degree in college (2-year)
            \item Bachelor’s degree in college (4-year)
            \item Master’s degree
            \item Doctoral degree
            \item Professional degree (JD, MD)
            \item Other
        \end{enumerate} 
\end{enumerate}

\subsection{Facial Action Unit coding and head orientation} \label{facs}

The head orientations are described in Table \ref{tab:head-orientation}.

\begin{longtable}{|p{3cm}|p{3cm}|}
\hline
\textbf{Head Orientation} & \textbf{Description}\\
\hline
pose\_Rx & Head Pitch \\
\hline
pose\_Ry & Head Yaw \\
\hline
pose\_Rz & Head Roll \\
\hline
\caption{Head Orientation coding}
\label{tab:head-orientation}
\end{longtable}

The coding system for facial action unit (based on \cite{facscoding}) is shown in Table \ref{tab:facs}.

\begin{longtable}{|p{5cm}|p{6cm}|}
\hline
\textbf{Action Unit (AU)} & \textbf{Description} \\
\hline
1 & Inner Brow Raiser \\
\hline
2 & Outer Brow Raiser (unilateral, right side) \\
\hline
4 & Brow Lowerer  \\
\hline
5 & Upper Lid Raiser \\
\hline
6 & Cheek Raiser  \\
\hline
7 & Lid Tightener  \\
\hline
9 (also shows slight AU4 and AU10) & Nose Wrinkler  \\
\hline
10 (also shows slight AU25) & Upper Lip Raiser  \\
\hline
11 & Nasolabial Deepener  \\
\hline
12 & Lip Corner Puller  \\
\hline
13 & Cheek Puffer  \\
\hline
14 & Dimpler  \\
\hline
15 & Lip Corner Depressor  \\
\hline
16 (with AU25) & Lower Lip Depressor  \\
\hline
17 & Chin Raiser \\
\hline
18 (with slight AU22 and AU25) & Lip Puckerer \\
\hline
20 & Lip stretcher  \\
\hline
22 (with AU25) & Lip Funneler  \\
\hline
23 & Lip Tightener  \\
\hline
24 & Lip Pressor  \\
\hline
25 & Lips part\\
\hline
26 (with AU25) & Jaw Drop \\
\hline
27 & Mouth Stretch \\
\hline
28 (with AU26) & Lip Suck  \\
\hline
41 & Lid droop  \\
\hline
42 & Slit  \\
\hline
43 & Eyes Closed  \\
\hline
44 & Squint  \\
\hline
45 & Blink  \\
\hline
46 & Wink \\
\hline
\caption{Facial Action Unit Coding System}
\label{tab:facs}
\end{longtable}

\subsection{Collected Data Information} \label{data}

From the study, the data we collected is shown in Table \ref{tab:data-description}.

\begin{longtable}{|p{3cm}|p{10.5cm}|}
\hline
\textbf{Data Type} & \textbf{Description}\\
\hline
Participant video & Three cameras, namely wheel camera (frequency of 30Hz), left A-pillar camera (frequency of 30Hz), and rear mirror camera (frequency of 60Hz). Each participant had three runs, totaling of 9 videos each participant. Camera resolution is 1920x1080 \\
\hline
Simulation video & Each participant has three simulation videos, one for each run. The videos record the simulation screen. \\
\hline
System audio & Audio from the system (computer voice, simulation sound) \\
\hline
Participant audio & Audio from participants (voice commands) \\
\hline
Heart rate & From the fitness tracker \\
\hline
Self-reported Ratings & From Q1, Q2, and Q3 from section 3.3 \\
\hline
Survey data & NASA-TLX after each run of each participant, and Simulation Sickness questionnaire, one for each participant \\
\hline
Game Logs & Logs from the word search game on the tablet. The files contain all the game events, such as word found, new game, hint used, current score, etc. \\
\hline
Interaction Logs & Logs from the system. The files contain all the interaction events, such as computer starts speaking, button presses, music ends, camera starts recording, etc. \\
\hline
\caption{Data description from the study}
\label{tab:data-description}
\end{longtable}

\subsection{Post-hoc statistical differences between baseline, correct, and unexpected stimuli} \label{stimulus-types-post-hoc}

\begin{figure}[htp]
    \centering
    \includegraphics[width=0.7\linewidth]{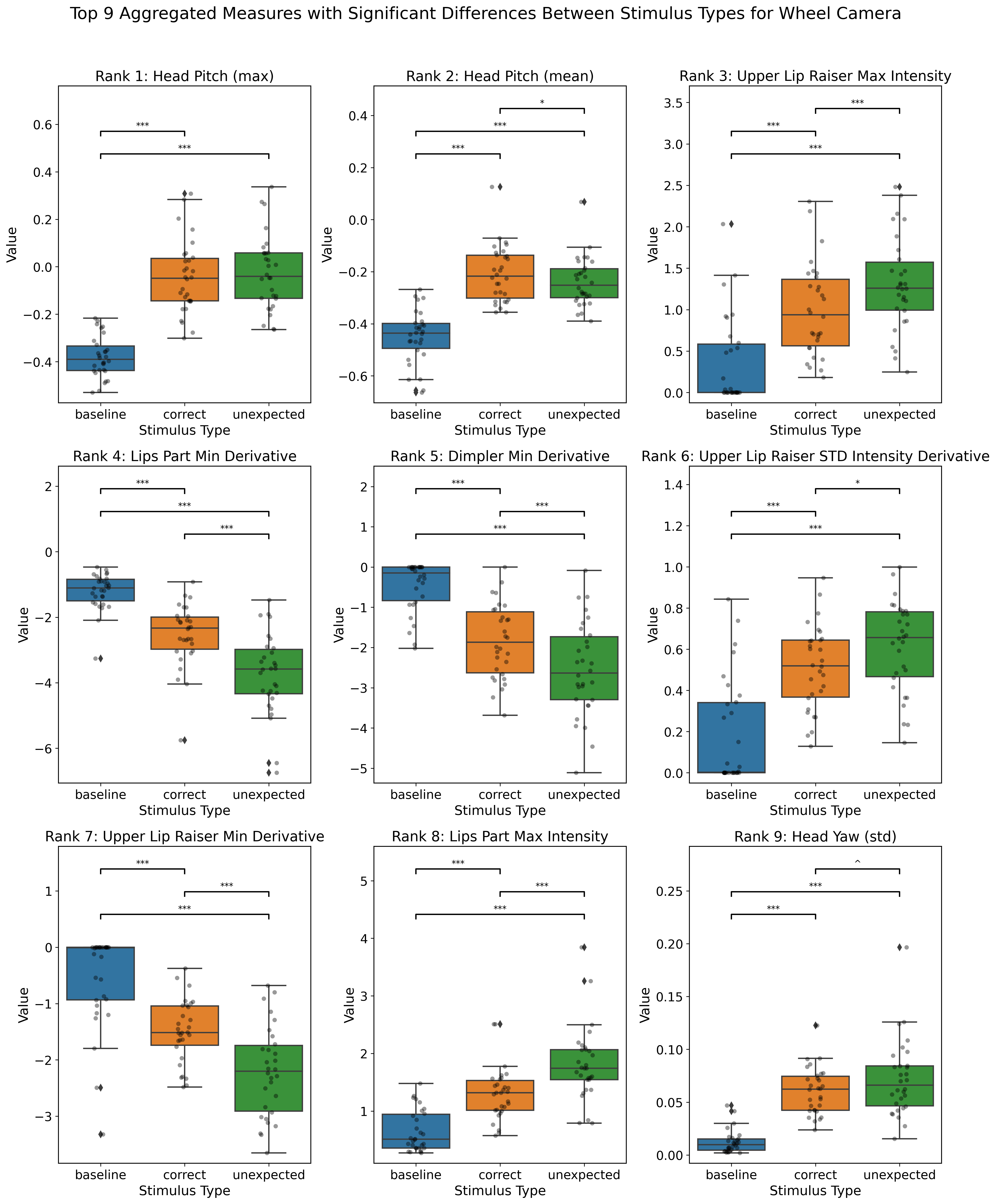}
    \caption{Post-hoc statistically significant differences among baseline, correct, and unexpected stimuli for the top 9 aggregated measures from the wheel camera. Black diamonds represent outliers. $\wedge$ represents $p\text{-value} < 0.05$, * represents $p\text{-value} < 0.01$, ** represents $p\text{-value} < 0.001$, and *** represents $p\text{-value} < 0.0001$.}
    \label{fig:baseline-camera-top-features}
\end{figure}

Figure \ref{fig:baseline-camera-top-features} shows the post-hoc statistical differences between the baseline, correct (iterations 1, 2, and 4), and unexpected stimuli (iteration 3) for the top 9 aggregated measures (e.g., maximum, minimum, mean, std) from the wheel camera.

\subsection{Viewpoints from the rear mirror and left A-pillar cameras}

\begin{figure}[htp]
    \centering
    \includegraphics[width=0.5\linewidth]{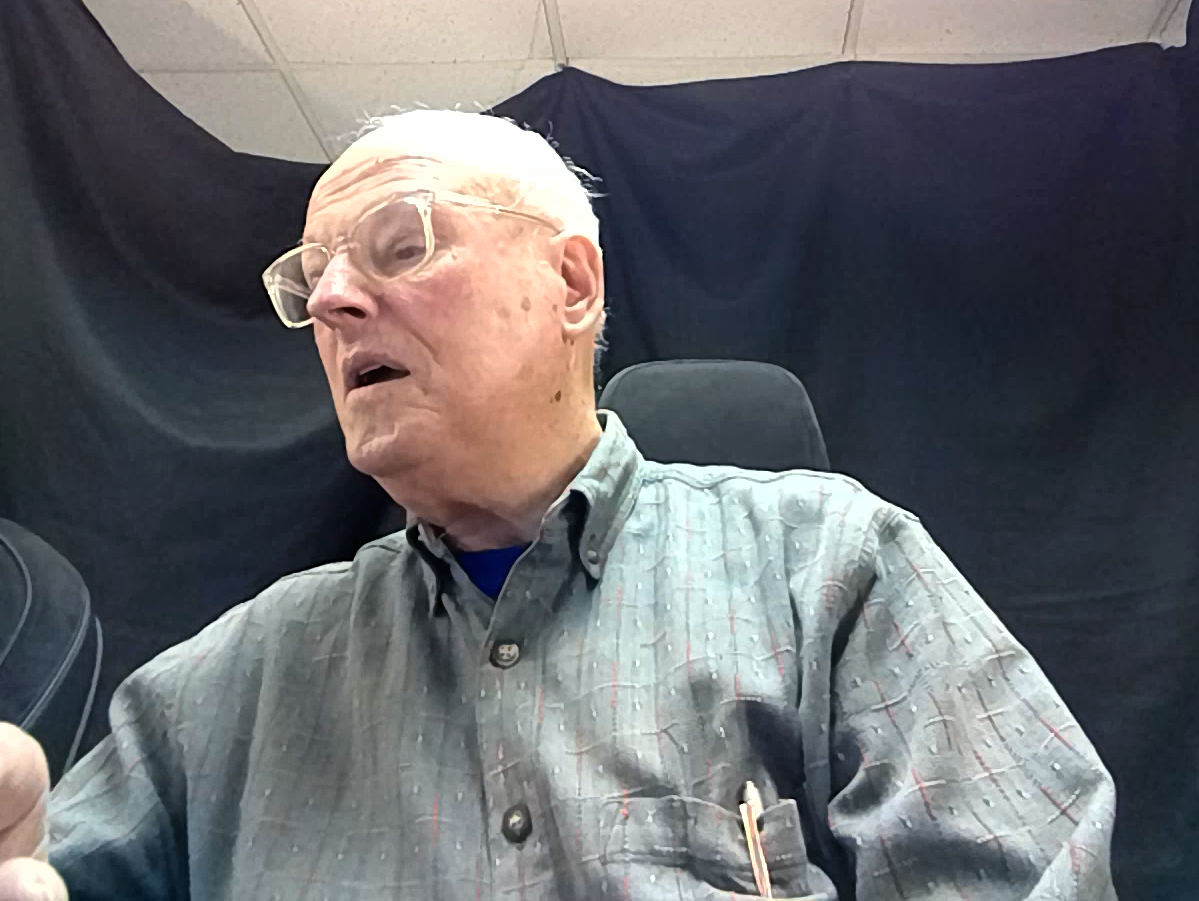}
    \includegraphics[width=0.5\linewidth]{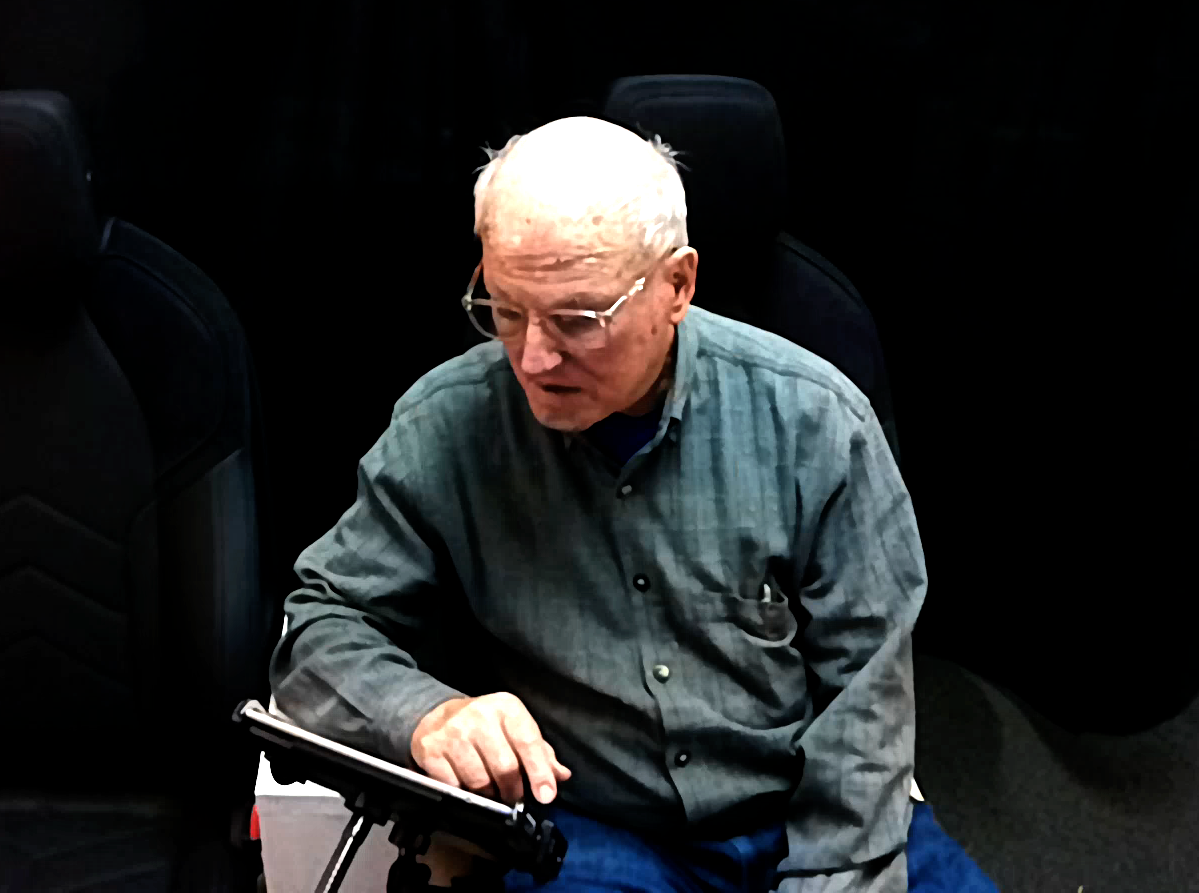}
    \includegraphics[width=0.5\linewidth]{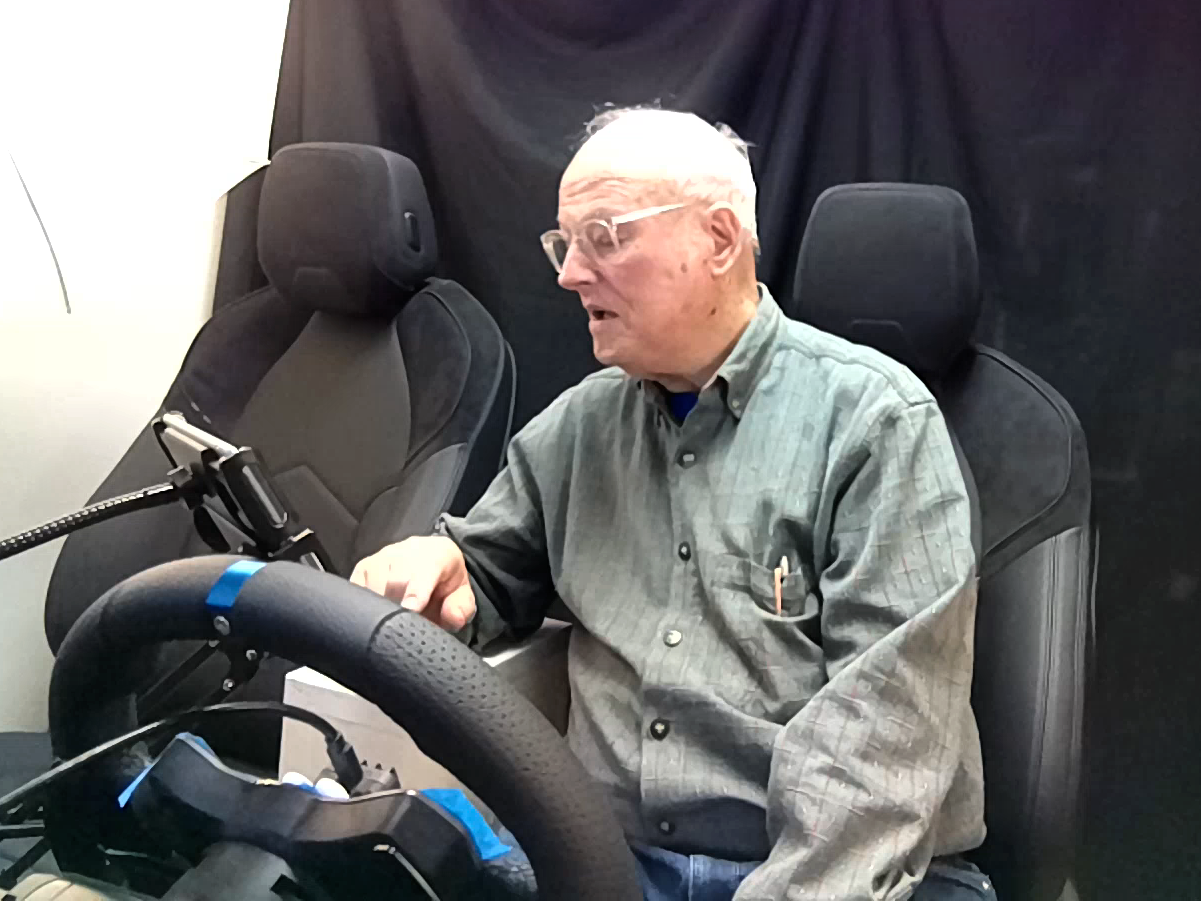}
    \caption{Viewpoints from the wheel camera (top), rear mirror camera (middle), and the left A-pillar camera (bottom).}
    \label{fig:other-viewpoints}
\end{figure}

The viewpoints from the rear mirror camera (middle) and the left A-pillar camera (bottom) compared with the viewpoint from the wheel camera (top) are shown in Figure \ref{fig:other-viewpoints}.

As we could observe from the viewpoints, the wheel camera provides the best angle for capturing clear facial expressions from participants as it almost angles directly to their face, from just a slight lower angle. The left camera misses half of participants' faces, and the lower portion of participants' faces might not always be entirely visible from the rear mirror camera. Thus, we only used data from the wheel camera for this analysis, and will analyze data from the other two cameras in our future work.

\end{document}

%% file: text/1_intro.tex
\section{Introduction}

As autonomous vehicles (AVs) become increasingly intelligent and capable of handling complex tasks, many manufacturers are introducing AI-powered personal assistants \cite{rege2024talking, wang2022conversational, abc2024aiagents, alibaba2023bmw, tesla2024grok} to enhance the user experience. These autonomous systems can potentially behave unexpectedly or make mistakes \cite{mueller2020humanlike}. Such behaviors can affect human perception of these systems and induce negative emotional responses and reactions within humans during interaction \cite{choi2021err}. However, unlike humans, autonomous vehicles are not always well-equipped to recognize and detect such negative emotional responses, especially when subtle or when the behaviors are considered non-critical. However, even non-critical unexpected behaviors can erode user confidence and trust, and it is important for systems, whether they are autonomous vehicles or robots, to recognize and mitigate these experiences through explanations \cite{ngo2024human} and other corrective behaviors \cite{graefe2022human}. Furthermore, negative human responses are a strong signal for refining machine learning models \cite{abdolmaleki2025learning, lin2020review, wang2023learning}.

In this work, we aim to address the problem of detecting different types of mismatches in expectation to system behaviors, as humans could react differently based on the type of expectation mismatch. 
Unlike most existing approaches that treat all unexpected system behaviors similarly, we recognize that users may perceive and respond differently based on the nature, severity, and context of the expectation mismatch. To our knowledge, currently available human emotion datasets in automotive settings focus on manual driving with general in-the-wild emotion detection \cite{xiao2022road, abdic2016driver, hori2016driver}, or in controlled robotic settings that focus primarily on binary detection of robot failures from user responses rather than the spectrum of unexpected behaviors \cite{bremers2023bystander, stiber2023using, gucsi2025hri}. Our contributed dataset bridges this gap by specifically examining fine-grained human emotional responses to various categories of unexpected system behaviors as triggering stimuli in an autonomous vehicle setting. We provide a preliminary analysis of three negative emotional responses, namely surprise, confusion, and frustration, which could be induced by unexpected behaviors of the system.

Unexpected system behaviors in fully automated vehicles create a mismatch between the driver's expectations and the reality of the situation. This discrepancy is a primary trigger for emotional and cognitive responses from the driver. The initial reaction to such unexpected events is often \emph{surprise}, according to the cognitive-evolutionary model of surprise \cite{reisenzein2019cognitive}. In an autonomous vehicle, an unexpected or unpredicted action by the AV can cause surprise \cite{wiegand2020d}. This initial surprise reflects the driver's awareness that the vehicle is not behaving as anticipated. If the reason for the unexpected behavior is not immediately apparent or easily understood, it can lead to \emph{confusion} \cite{wiegand2020d, koo2015did}. Confusion signals a lack of comprehension, often triggered by encountering something new and complex that is difficult to understand \cite{silvia2009looking}. When unexpected behavior leads to negative outcomes or impedes the driver's goals, it can lead to \emph{frustration}. Frustration is a negative emotional state resulting from perceived obstacles or unmet expectations \cite{berkowitz1989frustration}. This is particularly prominent when the system consistently underperforms and does not meet user expectations \cite{weidemann2021role}.

These emotional responses can be precursors to more severe user reactions, such as anger, road rage, or aggression \cite{ihme2018frustration}. Furthermore, experiencing surprise, confusion, or frustration due to unexpected behavior directly affects usability and adoption of these systems \cite{torggler2022beyond, hewitt2019assessing, koo2015did}. Differentiating between these states is crucial because each state may require a different kind of response from the vehicle.

Moreover, in the automotive domain, we consider situations in which unexpected behaviors occur inside the vehicle instead of outside on the road. Numerous studies have been performed on the detection and analysis of human reactions and emotional responses to events that occur on the road while driving \cite{zepf2019towards, li2021cogemonet, xiao2022road}. These events are related to safety and could potentially trigger high arousal responses in humans, even in studies conducted in a driving simulator. In this work, we aim to investigate in-vehicle stimuli that could potentially induce a more nuanced response from humans, which are innately harder to detect. In addition, in our study, participants were exposed to stimuli while in an autonomous vehicle and performing a secondary task instead of focusing on driving.

We conducted a human study within subjects ($N = 30$) with a driving simulator that runs a Unity-based simulation adapted from the StrangeLand simulator \cite{goedicke2022strangers}. We crafted three interaction scenarios with the in-car agent, namely playing music, ordering food, and visual alerts. To elicit the intended user responses of surprise, confusion, and frustration in each of these scenarios, we made the system behave unexpectedly in different ways, for example, playing the wrong music, messing up the food order or adding unexpected charges, and showing ambiguous alerts. We then performed a stimulus check with self-reported levels of unexpectedness, confusion, and frustration from participants to validate our stimuli, methods, and dataset. In addition, we analyze participants' facial expressions and head pose using OpenFace \cite{baltruvsaitis2016openface} to confirm the consistency of self-reported ratings and actual expressions while interacting with the systems.

Our paper makes the following contributions:

\begin{enumerate}
    \item A validated multi-modal multi-camera dataset for subtle human emotional responses and reactions to various unexpected system behaviors in a vehicle environment.
    \item Data analyses and evidence to differentiate emotional responses to different unexpected system behaviors.
\end{enumerate} 

%% file: text/2_background.tex
\section{Related work}

\subsection{Emotion detection in vehicles}

In the automotive domain, research on driver emotion recognition has primarily focused on Ekman's six basic emotions (happiness, sadness, disgust, anger, fear, and surprise) or its subsets, with anger and happiness being studied most frequently \cite{zepf2020driver, xiao2022road}. 
Studies have also used physiological signals like heart rate and electrodermal activity to detect stress \cite{rigas2011real, saeed2017personalized}. 
This prior research highlights the importance of driver emotional state detection as they significantly impact driving safety and performance, particularly in manually-driven vehicles \cite{wang2021survey, hori2016driver, abdic2016driver}. However, as automation increases in modern vehicles, our work extends beyond simply detecting emotions to understanding the causal relationship of how specific in-vehicle events trigger emotional responses. 

\subsection{Detecting User Responses to Autonomous Systems}

Recent research examines emotional states in autonomous driving contexts, where they influence trust, takeover performance, and user experience \cite{karas2023audiovisual, sanghavi2020effects, braun2021affective, wu2021designing}. Lee et al. \cite{lee2021human} designed a human–machine interface (HMI) for semi-autonomous vehicles that processes driver emotions using bio-signals, such as Galvanic Skin Response (GSR) and Photoplethysmography (PPG) for vehicle handovers. Similarly, Ling et al. \cite{ling2021towards} developed an emotionally adaptive AV framework that uses Electroencephalogram (EEG) sensors to infer driver preferences from emotional states to correspondingly adapt vehicle behavior in a driving simulator. Braun et al. \cite{braun2021affective} reported that negative emotions could be triggered by unsatisfactory interaction with the user interface of the car, as well as insufficient capabilities of the vehicle systems. Despite growing research in this field, relatively few studies have examined how users emotionally respond when non-critical systems (such as in-vehicle entertainment or personal assistants) behave unexpectedly. However, as in-cabin monitoring becomes more sophisticated, we anticipate a shift toward better understanding passenger affect in highly automated vehicles \cite{karas2023audiovisual}.

More broadly, multiple studies have examined user reactions to the performance and failures of autonomous systems through various data sets \cite{candon2024react, bremers2023bystander, adegun2020facial, bremers2024social, stiber2023using}. The BAD (Bystander Affect Detection) dataset \cite{bremers2023bystander} captures bystander reactions (facial expressions and body movements) to videos of human and robot failures through webcam recordings. It aims to enable robots to detect errors by observing implicit reactions from bystanders. However, this dataset involves passive observation of robot failures rather than direct interaction. On the other hand, the REACT database tracks human reactions and evaluative feedback to robots during a collaborative photography task \cite{candon2024react}. Similarly, HRI-SENSE is a multimodal dataset that examines user social, physical, and emotional responses to robot behaviors during a ``Burger Assembly'' task \cite{gucsi2025hri}. The database from \cite{stiber2023using} focuses on facial responses to robot errors during physical human-robot interactions. While these existing datasets provide valuable insights into human reactions to robot failures, they do not focus on robot behaviors based on their impact on user reactions.
While existing datasets effectively capture human reactions to robot failures, potentially enabling robots to recognize their own errors and assess their performance, they don't provide insights into how robots could adapt their behaviors in response to user emotional feedback.

\subsection{Detecting Surprise, Confusion, and Frustration}

Emotion detection research in the automotive domain has explored various approaches for identifying surprise, confusion, and frustration. Zepf et al. \cite{zepf2019towards} conducted an experiment to induce and detect frustration while driving in various ambient traffic, along with time pressure. They added a manipulated secondary task of interacting with a speech interface to induce further frustration. Similarly, in the driving simulator study by Ihme et al. \cite{ihme2018frustration}, participants were induced with frustration with a parcel delivery assignment where the road events (traffic, other car, red lights, etc) contributed to such frustration. They manually coded facial action units (AUs) \cite{ekman1978facial} and identified AUs in the mouth area, such as AU23 (lip tightener), AU24 (lip pressor), AU10 (upper lip raiser), AU12 (lip corner puller), AU17 (chin raiser), and AU20 (lip strecher), to correlate with frustration. 

For surprise detection, Li et al. \cite{li2021cogemonet}  developed methods incorporating both temporal facial expressions and cognitive characteristics like age, gender, and driving experience to predict emotional states, including neutral and surprise, in a driving simulator. However, this work focused on out-of-vehicle safety-critical aspects.  For confusion, researchers such as Ucar et al. \cite{ucar2023driver} have extracted vehicle trajectory data and identified turn loops in traffic that are tagged as confusion zones to provide driving guidance in those zones. 

\subsection{Understanding user state via computer vision and multimodal data}


Many researchers have found that multimodal approaches combining several data sources such as facial and head signals (e.g., facial action units), biophysiological signals (i.e., heart rate variability, galvanic skin response), speech (i.e., speech content) and driver behavior (i.e., changes in steering wheel and pedal activations) typically yield superior performance in emotion recognition \cite{zepf2020driver}. Hori et al. \cite{hori2016driver} used multi-modal features such as driving condition (i.e., driving action, steering angle), traffic condition (i.e., road type, traffic signals), driver's behavior (i.e., voice activity,  gaze direction), and location information (i.e., distance to the goal) to classify driver confusion in real driving. In their work, LSTM RNN outperformed other classification networks due to being able to make use of temporal context. For frustration while interacting with voice-based navigation, Abdic et al. \cite{abdic2016driver} used both video and audio streams of the interaction to yield a high-frequency of frustration detection. They state that additional data streams, such as heart rate and skin conductance, is also important to detect frustration.

Beyond automotive applications, emotion detection has been studied in other domains. Reisenzein et al. \cite{reisenzein2019cognitive} argued that surprise is evoked by unexpected events, and followed by other physiological responses such as increased skin conductance, heart rate changes, and pupil dilation. Adegun et al. \cite{adegun2020facial} show that it is possible to detect micro-expression using temporal feature extraction and a machine learning algorithm with fast learning speed. Borges et al. \cite{borges2019classifying} considered Facial Action Coding schema to train an LSTM neural network to detect instances of confusion expressed in a social context (e.g., map direction task). The results showed that certain facial action units related to mouth expressions (AU25, AU26, and AU27) were the most important contributing factors to the model's performance. 

Understanding a driver's availability for interaction is a crucial aspect of their state, as highlighted by Semmens et al. \cite{semmens2019now}. Their study found that drivers prefer to be spoken to when driving straight at a constant speed or when stopped, and they dislike interruptions during driving maneuvers or when off-course, from the collected video data of drivers and the vehicle data. Complementing this, Graefe et al. \cite{graefe2022human} explore user-adaptivity in intelligent vehicles, where systems proactively change behavior based on user characteristics and preferences. These works emphasize the necessity of a human-centered approach to understanding and responding to the user's state in automotive HMI design.


Despite these advances, significant research gaps remain in emotion detection studies for automotive contexts. Most research has focused on driver emotions to external driving conditions than user reactions to in-vehicle systems, with limited investigation into emotional responses to unexpected behaviors from autonomous vehicles or intelligent personal agents. Importantly, existing studies fail to systematically categorize different types of user responses to unexpected system behavior based on the nature, severity, and context of the unexpected behavior. Furthermore, most studies have focused on manually driven vehicles rather than autonomous behaviors, with less emphasis on user experience with common non-critical interactions. While some studies have explored reaction detection in robotics contexts (e.g., BAD, REACT), they primarily focus on robot failures during passive observation or specific collaborative tasks, not on in-vehicle interactions.  While multimodal data and approaches show improved performance in emotion recognition, research combining multiple data streams for detecting and differentiating between various categories of user responses to unexpected autonomous vehicle behaviors remains underdeveloped.

%% file: text/4_study.tex
\section{Human Study}

To address the lack of research on how humans respond to unexpected behaviors from in-vehicle systems in autonomous vehicles, we conducted a within subjects study examining users' emotional reactions to various unexpected in-vehicle behavior. The study was approved by our institution's Institutional Review Board.

\subsection{Triggered Response Description}
We focus on stimuli to evoke three specific responses, namely, \textit{surprise}, \textit{confusion}, and \textit{frustration} based on their corresponding characteristics described in \cite{reisenzein2019cognitive, silvia2010confusion, zepf2019towards}. Thus, we expected to see the responses as follows:

1. \textit{Surprise} is described as a response with a high level of unexpectedness. However, the duration of the response is short, as the reasons for it either become clear on its own or are immediately followed by the stimulus. Thus, \textit{surprise} response is also described with a lower level of confusion and frustration.

2. \textit{Confusion} is described as a response with a high level of unexpectedness, but without any obvious reasons or the reasons are unclear and ambiguous, unlike \textit{surprise}. In addition, \textit{confusion} could be followed by a mild feeling of frustration due to lack of understanding.

3. \textit{Frustration} is described as a response with a high level of unexpectedness, without obvious reasons, and actively blocking the goals of the participants. The blocking of one's goal could also be repeated or have a longer duration.

\subsection{Scenario and Stimuli Description}
To investigate the effect of the stimuli, we crafted three scenarios that contained non-critical in-vehicle events (e.g., stimuli, more details in Appendix) that triggered the desired emotional responses in participants.

1. \textit{Music Playing}: The system played songs requested by the participants using voice commands (correct stimuli). 
To induce \textit{surprise}, the system played songs from an artist who had a very similar name to the requested artist (unexpected stimuli). 
To induce \textit{confusion}, the system played songs from an artist that are completely irrelevant to the requested artist (unexpected stimuli). 
To induce \textit{frustration}, the system refused to play songs by the requested artist and repeatedly tuned to radio channels that do not play songs, such as the weather channel (unexpected stimuli). 

2. \textit{Food Ordering}: Participants used voice commands to get the system to order food from a local restaurant (correct stimuli).
To induce \textit{surprise}, the system ordered food but with an unexpected additional fee for delivery (unexpected stimuli). 
To induce \textit{confusion}, the system ordered the wrong food (unexpected stimuli). 
To induce \textit{frustration}, the system kept failing to place an order and repeatedly asked participants to try again a few times (unexpected stimuli).

3. \textit{Alert}: The system displayed visual and audio alerts to participants about the simulated autonomous driving experience (correct stimuli).
To induce \textit{surprise}, the system showed an alert preemptively, and the reason soon became clear (unexpected stimuli). 
To induce \textit{confusion}, the system showed an alert of an object that was not in the scene for no particular reason (unexpected stimuli). 
To induce \textit{frustration}, the system showed a wrong alert, accompanied by a continuously repeated voice notification that could not be turned off (unexpected stimuli). 

\subsection{Study Procedure}

The study was conducted in-person in our institution's lab space. Participants were recruited through a city-wide Web platform, flyers, and word of mouth.

We recruited $N = 30$ participants in total from varying races and ethnicities (white, asian, middle eastern) and educational backgrounds (doctorate, college degree, high school, etc.). 40\% of the participants were male. The mean age of participants was 31.4 (SD=17.1), with a broad age range of 18 to 84 years old. In the pre-study survey, we asked four 7-point Likert-scale questions about their familiarity with technology, experience with video games, familiarity with autonomous vehicles, and familiarity with voice agents. In general, the participants were familiar with technology (93.3\%  participants rated 5 (Moderately Agree) or higher), had experience with video games (73.3\% participants rated 5 or higher), were familiar with autonomous vehicles (46.7\% participants rated 5 or higher) and were familiar with voice agents (90\% participants rated 5 or higher).



The study began with participants being briefed about the study, signing the consent form, and filling out a pre-study survey about their familiarity with technologies, video games, autonomous vehicles, and voice-based agents. The participants were then introduced to the driving simulator and how to interact with the system with voice and buttons. Participants were asked to engage in a secondary task of playing a word search game on a tablet. During the game, participants are exposed to the stimuli within the scenarios by interrupting task prompts that appear on the tablet that block them from playing until the tasks are completed. 
After every planned interaction with the system, participants were asked to complete a 3-question survey on the 7-point Likert scale on the tablet about how they felt about the system behavior, with ``1'' being ``Not at all'' and ``7'' being ``Definitely''. The survey questions are as follows:

Q1. Was the system behavior unexpected?

Q2. Was the system behavior confusing or unclear?

Q3. Was the system behavior frustrating?

Following Section 3.1, we expected that a \textit{surprise} response would have a high rating on Q1 only, a \textit{confusion} response would have high ratings on Q1 and Q2, and a \textit{frustration} response would have high ratings on all three questions.

Each participant experienced a practice run to become familiarized with the system, followed by three experiment runs, each corresponding to a specific response and having all three interaction scenarios. For each scenario inside a run, we presented a sequence of four stimuli through four iterations. The first two were ``correct'' stimuli as the system followed participants' requests. The third one was an ``unexpected'' stimulus as the system presented an unexpected behavior corresponding to the intended response. The final stimulus was a ``recover'' stimulus where the system responded correctly to the participants' requests. To eliminate order effects, we counterbalanced the three responses and three scenarios over the three runs with orthogonal Latin Squares \cite{grant1948latin}. The use of different Latin Squares also reduced bias by varying the scenarios across the responses.

After each run, participants completed the NASA-TLX survey \cite{hart1988development} to assess cognitive workload during the run. After all three runs, participants completed a demographic survey, a simulation sickness questionnaire \cite{kennedy1993simulator}, followed by a short interview to inquire into their perceptions of the system behavior and their desired communication methods from the system after unexpected events. The participants were then compensated for their time and left the study site.

\subsection{Data Collection}

During the study, all data was collected using a driving simulator. The setup of the simulator is shown in Figure \ref{fig:teaser}. 
Participants sat in the driver's seat and played a word search game on the tablet. In our dataset, the tablet was handheld for the first 15 participants and mounted on the stalk for the remaining 15 participants.
This is to simulate real-life scenarios where people could use their own handheld devices (i.e., phone, game devices) or use the vehicle's dashboard (i.e., infotainment dashboard) during the ride.
Occasionally, participants interacted with the vehicle system using voice or physical buttons on the wheel. The vehicle simulation was based on an existing Unity-based driving simulation called StrangeLand \cite{goedicke2022strangers} and was adapted for our study. During the study, three cameras recorded the interactions from different angles: a rear mirror camera, a wheel camera, and a left A-pillar camera. The Unity simulation screen was also recorded, along with the system voice and participants' voices in separate tracks. After every planned interaction, participants answered the previously described 3-question survey about the system's behavior. Additionally, participants wore a fitness watch that tracked their heart rate during interactions. Every vehicle system behavior, game tablet event, and participant interaction were logged with timestamps (i.e., word found, participants' requests, button press, etc.). 


A detailed description of the collected data is in our Appendix. In this paper, we present an analysis of a subset of the dataset, namely self-reported survey responses (stimulus check) and facial features from videos.

%% file: text/5_results.tex
\section{Results}
In this section, we describe the results of a stimulus check to confirm that our stimuli indeed induced the responses we intended and a post-hoc analysis of facial features to investigate participants' responses to the stimuli. 

\subsection{Self-reported Survey Answers as Stimulus Check}

The goals of the three-question stimulus check survey were two-fold. First, we wanted to confirm that participants responded differently to the ``correct'' and ``recover'' stimuli (e.g., iterations 1, 2, and 4) compared to the ``unexpected'' stimuli (e.g., iteration 3). Second, we wanted to confirm whether participants respond in a unique manner pertaining to each emotional response (e.g., surprise, confusion, frustration) triggered by unexpected stimuli.

\begin{figure}[htp]
    \centering
    \includegraphics[width=\linewidth]{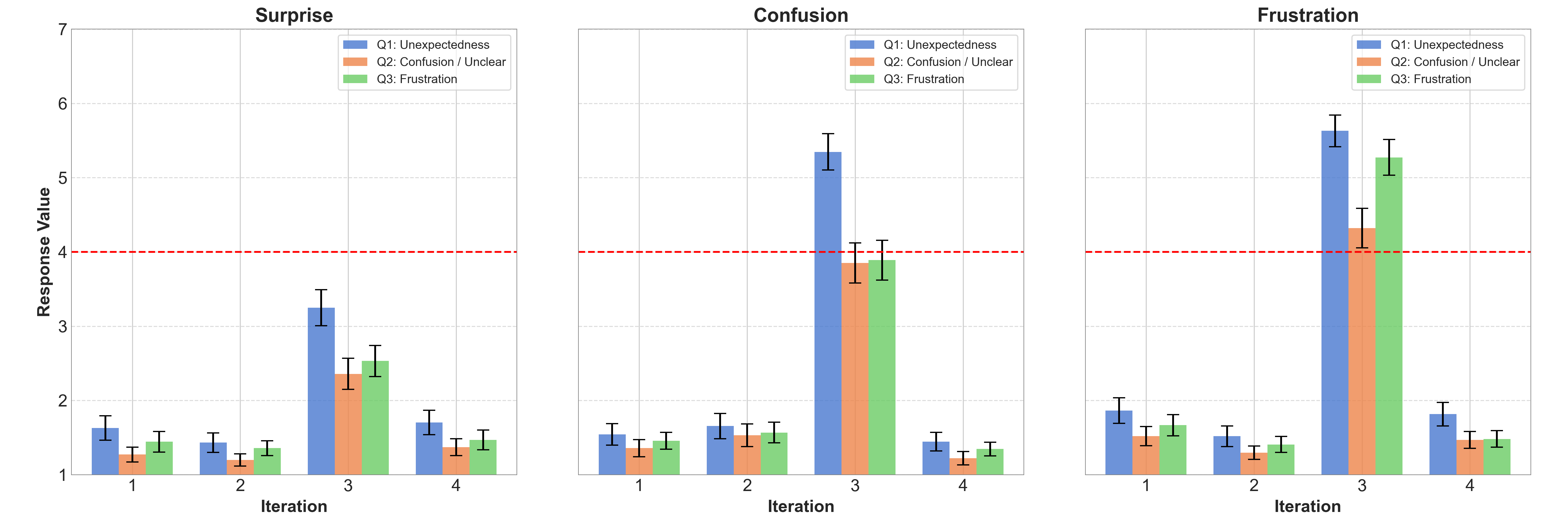}
    \caption{Ratings of participants in three triggered responses on all iterations (error bars are standard error). The mid-point of the survey scales is shown as a dotted red line for easier reference.}
    \label{fig:stimulus_check}
\end{figure}

The results of the stimulus check among participants are shown in Figure~\ref{fig:stimulus_check}. We excluded the first three participants from this analysis since we slightly modified the survey for clarity.

Using repeated measures one-way ANOVAs, we found statistically significant differences between correct / recovery stimuli (e.g., iterations 1, 2, and 4) and unexpected stimuli (e.g., iteration 3) for Q1 ($F(1, 26) = 283.81, p < 0.0001$), Q2 ($F(1, 26) = 84.54, p < 0.0001$), and Q3 ($F(2, 46) = 125.11, p < 0.0001$). Participants noticed and perceived unexpected system behaviors differently from other behaviors.
The confusion response had a higher level of unexpectedness (Q1), confusion (Q2), and frustration (Q3) than surprise. 
In the frustration response, participants had similar levels of unexpectedness (Q1) and confusion (Q2) but with increased frustration (Q3).

In addition to the survey response trends, we performed repeated measures ANOVAs to compare the effect of the three triggered responses on participants' ratings of Q1, Q2, and Q3. 

For Q1, there was a statistically significant difference between the triggered responses ($F(2, 46) = 37.10, p < 0.0001$). A post-hoc paired t-test with Bonferroni correction \cite{weisstein2004bonferroni} revealed that the mean Q1 responses of \textit{confusion} and \textit{frustration} are significantly higher than those of \textit{surprise} ($\mu = 6.58, p-corrected < 0.0001$ and $\mu = 7.59, p-corrected < 0.0001$, respectively). There was no significant difference between \textit{confusion} and \textit{frustration}.

For Q2, there was also a statistically significant difference between the triggered responses ($F(2, 46) = 24.62, p < 0.0001$). A post-hoc paired t-test with Bonferroni correction revealed that the mean Q2 responses of \textit{confusion} and \textit{frustration} are significantly higher than those of \textit{surprise} ($\mu = 6.02, p-corrected < 0.0001$ and $\mu = 7.35, p-corrected < 0.0001$, respectively). Once again, there was no significant difference between \textit{confusion} and \textit{frustration}.

For Q3, there was also a statistically significant difference between the triggered responses ($F(2, 46) = 45.75, p < 0.0001$). A post-hoc paired t-test with Bonferroni correction revealed that the mean Q3 responses of \textit{confusion} and \textit{frustration} are significantly higher than those of \textit{surprise} ($\mu = 4.05, p-corrected < 0.001$ and $\mu = 9.98, p-corrected < 0.0001$, respectively). This time, the mean Q3 response of \textit{frustration} is significantly higher than that of \textit{confusion} ($\mu = 5.72, p-corrected < 0.0001$).

The results of the repeated measures ANOVA show that the participants' frustration levels (Q3) for the three triggered responses were significantly different. The responses for unexpectedness (Q1) and confusion (Q2) were significantly different between surprise and confusion/frustration but not between confusion and frustration themselves. Although excluded from this analysis, participants 1, 2, and 3 had the same stimuli as the rest of the participants. Since we confirmed the stimulus check from the other 27 participants, we assume that the first three participants perceived the system behaviors in a similar way. Thus, we included those three in the facial features data analysis.

\subsection{Facial Features Data from Videos}

In this paper, we focus on the facial features extracted from videos of participants from the wheel camera. This viewpoint represented the best possible angle for facial features using OpenFace \cite{baltruvsaitis2016openface}. 
Other viewpoints from the rear mirror camera and the left A-pillar camera are not optimal, as portions of participants' faces from those viewpoints are somewhat not visible, thus creating less reliable facial features.
We specifically investigated the snippets of interactions (e.g., iterations 1, 2, 3, and 4) and baseline (collected by having participants sit still, look at the simulation screen, and do nothing at the beginning of the study). We focus on the Facial Action Units (AUs) intensities and head rotations values (roll, pitch, yaw) provided by OpenFace due to their proven effectiveness in recognizing emotions \cite{zepf2019towards, ihme2018frustration}. 

\begin{figure}[htp]
    \centering
    \includegraphics[trim={0.3cm 0.2cm 0cm 0cm}, clip, width=0.7\linewidth]{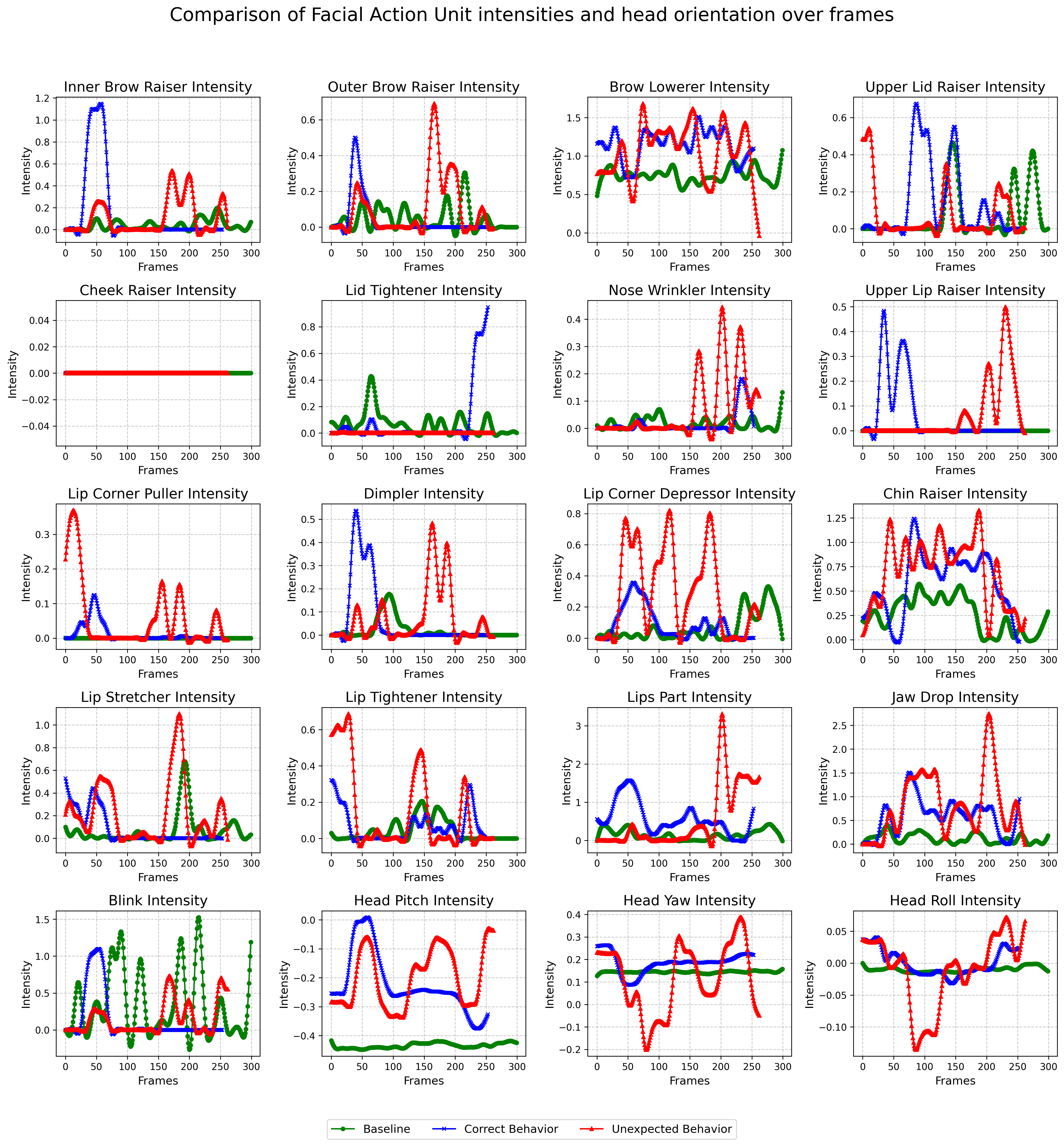}
    \caption{Example of raw values AU intensities and head orientation values over frames between baseline (green), correct (blue), and unexpected (red) stimuli in one participant (P8) from the wheel camera.}
    \label{fig:AU-headpose-raw-comparison}
\end{figure}

\begin{figure}[htp]
    \centering
    \includegraphics[width=0.7\linewidth]{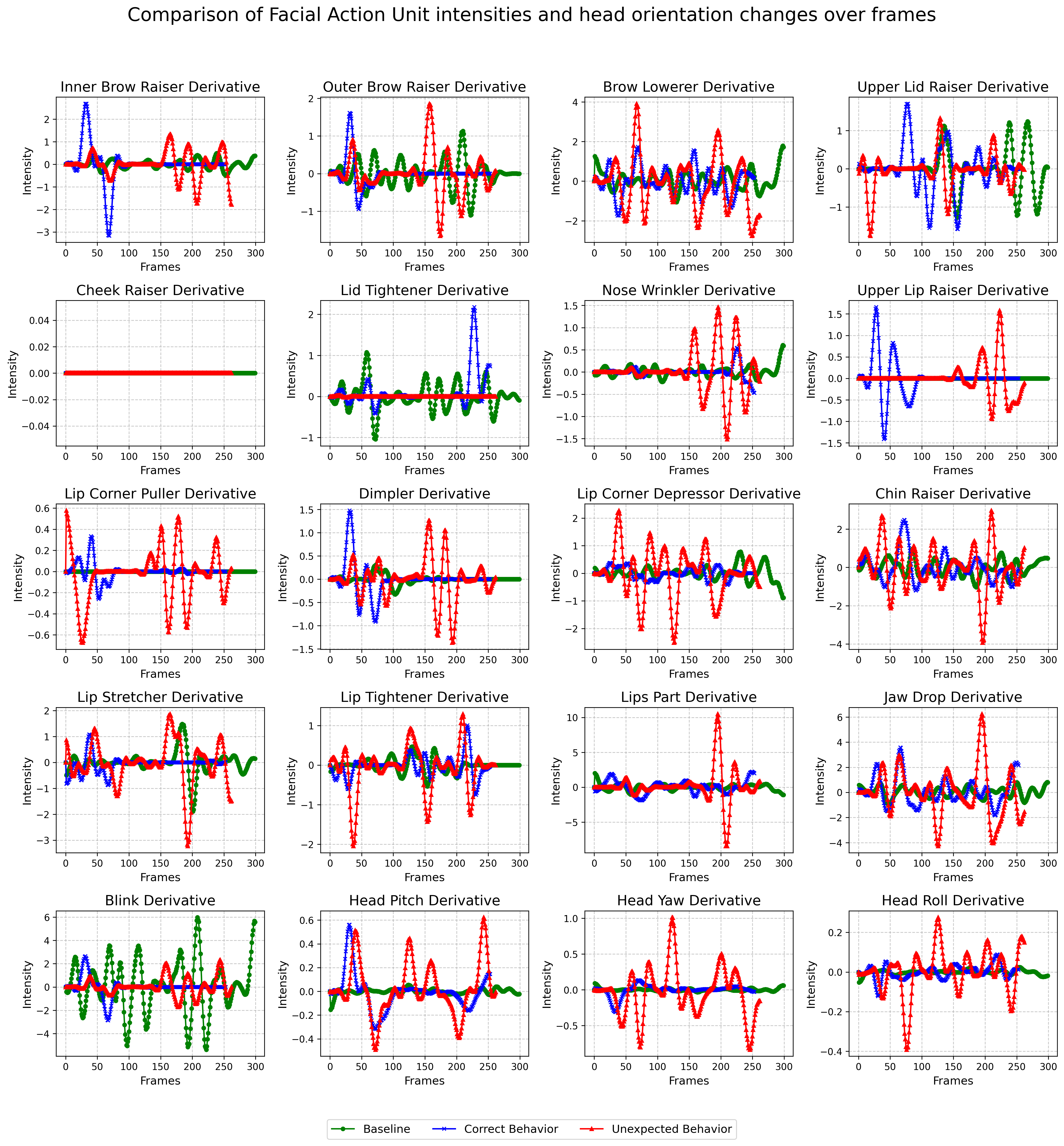}
    \caption{Example of derivatives of AU intensities and head orientation values over frames between baseline (green), correct (blue), and unexpected (red) stimuli in one participant (P8) from the wheel camera.}
    \label{fig:AU-headpose-derivative-comparison}
\vspace{-3mm}
\end{figure}

First, we applied a median filter (with a window size of 5 frames) followed by a fourth-order Butterworth low-pass filter (with a cutoff frequency of 0.1 Hz). Such filtering helped with denoising and removing spikes in the data. For each AU and head rotation orientation, we calculated the first derivative of such values. An example of the raw and derivative values of the AU intensities and head orientation from the wheel camera is shown in Figures~\ref{fig:AU-headpose-raw-comparison} and \ref{fig:AU-headpose-derivative-comparison}, respectively.

As can be seen in the example in Figures~\ref{fig:AU-headpose-raw-comparison} and \ref{fig:AU-headpose-derivative-comparison}, we observed some interesting distinctions from the facial expressions of the participants when the system behaved correctly (blue) and unexpectedly (red). For example, there are higher peaks in AU25 (Lips Part), AU26 (Jaw Drop), AU15 (Lip Corner Depressor), AU09 (Nose Wrinkler), pose\_Ry (Head Yaw), and in their derivatives. Moreover, we also observed differences from baseline facial features (green) compared to correct (blue) and unexpected (red) stimuli from AU14 (Dimpler), AU25 (Lips Part), AU26 (Jaw Drop), pose\_Rx (Head Pitch), pose\_Ry (Head Yaw), and pose\_Rz (Head Roll), etc.

\begin{figure}[htp]
    \centering
    \includegraphics[width=0.7\linewidth]{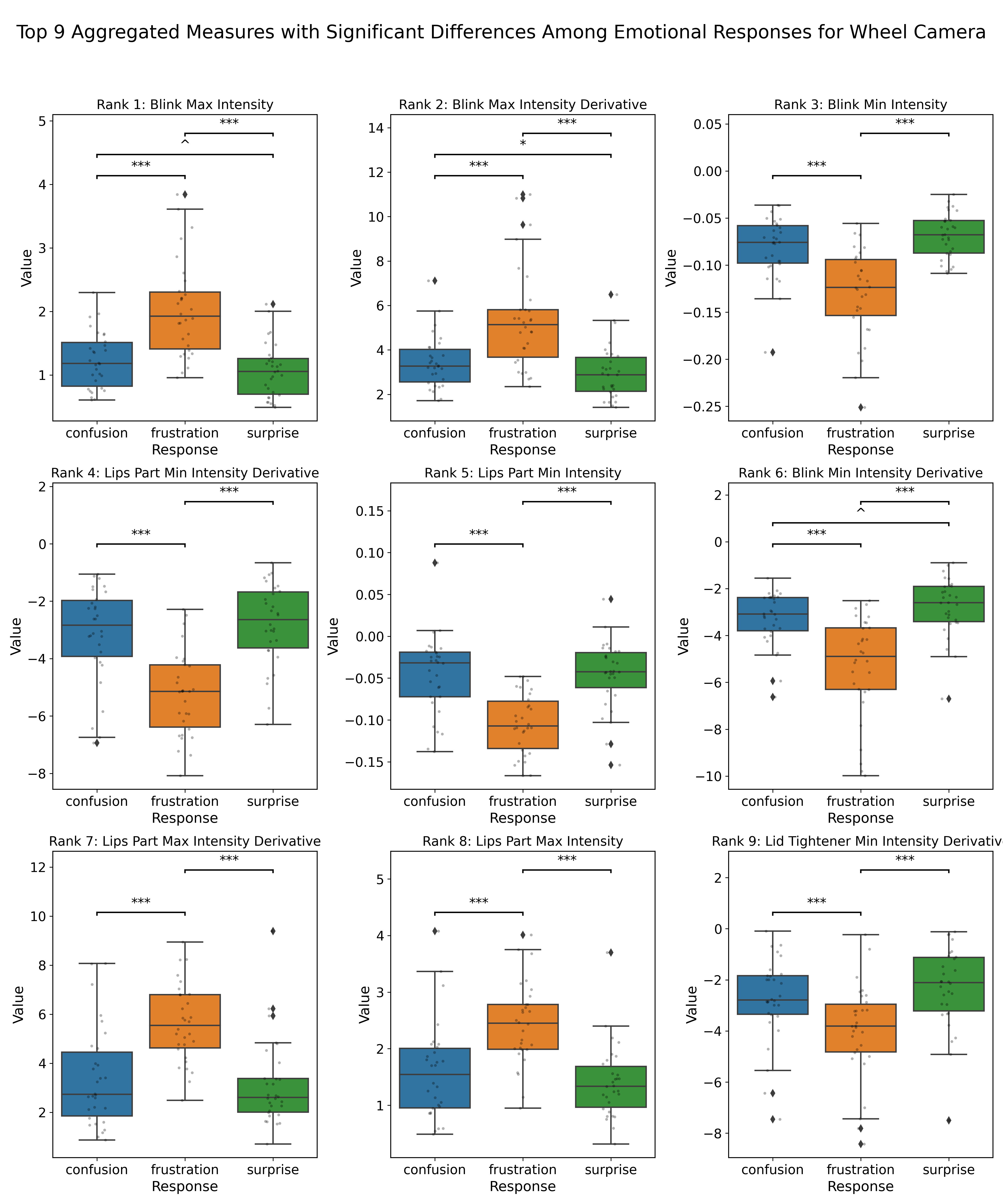}
    \caption{Post-hoc statistically significant differences among emotional responses for the top 9 aggregated measures from the wheel camera. Black diamonds represent outliers. $\wedge$ represents $p\text{-value} < 0.05$, * represents $p\text{-value} < 0.01$, ** represents $p\text{-value} < 0.001$, and *** represents $p\text{-value} < 0.0001$.}
    \label{fig:camera-top-features}
\vspace{-3mm}
\end{figure}

We investigated the characteristics of AUs and head orientations by calculating their maximum, minimum, mean, and standard deviation values (e.g., aggregated measures) of the raw and derivative values for each interaction window of each participant. We performed one-way repeated measures ANOVAs to see whether there were significant differences between baseline, correct stimuli (iterations 1, 2, 4), and unexpected stimuli (iteration 3). Moreover, we performed one-way repeated measures ANOVAs to see whether there were differences between the three triggered emotional responses in unexpected stimuli. The top 10 most significant features (with effect size > 0.5) for both are shown in Table~\ref{tab:openface_baseline_features}.
Please note that Table \ref{tab:openface_baseline_features} ranks the top 10 features (e.g., AUs and head orientations) by the number of significant aggregated measures (e.g., min, max, mean, std), while Figure
\ref{fig:camera-top-features} shows the post-hoc results of the aggregated measures.

\begin{table}[h!]
\centering
\begin{tabular}{ | m{3em} | m{6cm}| m{6cm} | } 
  \hline
  \textbf{Rank} & \textbf{Significant features for different stimulus types} & \textbf{Significant features for different emotional responses} \\ 
  \hline
  1 & AU25 (Lips Part)    & AU45 (Blink)\\ 
  \hline
  2 & AU10 (Upper Lip Raiser) & AU14 (Dimpler)  \\ 
  \hline
  3 & AU14 (Dimpler)  & AU25 (Lips Part)  \\ 
  \hline
  4 & AU23 (Lip Tightener) & AU07 (Lid Tightener) \\ 
  \hline
  5 & pose\_Ry (Head Yaw)  & AU12 (Lip Corner Puller)  \\ 
  \hline
  6 & AU04 (Brow Lowerer)  & AU26 (Jaw Drop)  \\ 
  \hline
  7 & AU01 (Inner Brow Raiser) & AU23 (Lip Tightener) \\ 
  \hline
  8 & AU17 (Chin Raiser)  & AU10 (Upper Lip Raiser)  \\ 
  \hline
  9 & pose\_Rx (Head Pitch) & AU04 (Brow Lowerer)  \\ 
  \hline
  10 & AU12 (Lip Corner Puller) & AU17 (Chin Raiser)  \\ 
  \hline
\end{tabular}
\caption{Top 10 features that had significant differences between baseline, correct, and unexpected stimuli across all responses (Second column) and top 10 features that had significant differences between the three responses for unexpected stimuli (Third column).}
\label{tab:openface_baseline_features}
\vspace{-7mm}
\end{table}


We also performed post-hoc t-tests with Bonferroni correction for the above two analyses for the statistically significant aggregated measures from the repeated measures ANOVA. 
For different stimulus types, we observed a statistically significant difference among the baseline, correct, and unexpected stimuli across multiple human behavior aggregated measures.
For different emotional responses, the results for the aggregated measures from the wheel camera are shown in 
Figure~\ref{fig:camera-top-features}.
There are multiple statistically significant differences between \textit{surprise} and \textit{frustration}, as well as between \textit{confusion} and \textit{frustration}. 
We also observed a statistically significant difference between \textit{surprise} and \textit{confusion} within human behavior aggregated measures such as AU45 (Blink) max intensity ($\mu = 0.16, p-corrected = 0.041$), AU45 max intensity derivative ($\mu = 0.47, p-corrected < 0.01$), and AU45 min intensity derivative ($\mu = 0.51, p-corrected = 0.037$). There were other significant differences in the top 9 not shown here, such as AU07 (Lid Tightener) max intensity derivative ($\mu = 0.53, p-corrected = 0.025$), AU10 (Upper Lip Raiser) max intensity derivative ($\mu = 0.38, p-corrected = 0.025$), AU14 (Dimpler) max intensity derivative ($\mu = 0.54, p-corrected = 0.038$), etc.

%% file: text/7_discussion.tex
\section{Discussion}

\subsection{Stimulus Check}

The results of our stimulus check, comparing participants' self-reported survey responses with the study stimuli, support our methods. The findings showed statistically significant differences between \textit{surprise} and \textit{confusion} as well as between \textit{surprise} and \textit{frustration} on the levels of unexpectedness (Q1) and confusion (Q2). Moreover, we saw statistically significant differences between all three triggered responses on the level of frustration (Q3). 

The results indicate that the participants holistically perceived the three categories of unexpected behaviors distinctly. Surprise is described as an emotion that arises from schema disruption or unexpectedness \cite{reisenzein2019cognitive}, which we find as well. Silvia et al. \cite{silvia2010confusion} argued that confusion has a high level of unfamiliarity or unexpectedness and a low level of understanding. We find evidence for this from the significant differences in unexpectedness (Q1) and level of confusion (Q2) between the surprise and confusion responses. Frustration can be temporal and can be triggered through repeated unexpected interactions over a long period of time \cite{zepf2019towards, weidemann2021role}. In our scenarios with repeated unexpected system behavior, we see a notably higher level of frustration (Q3). We found that it was harder to distinguish between confusion and frustration, especially in Q1 and Q2. This could be because both confusion and frustration can have shared antecedents, such as the inability to comprehend the situation. This is commonly observed in the learning context \cite{liu2013sequences} that confusion and frustration are recently viewed as a joint construct called, ``confrustion''. However, in our study, we are able to distinguish them on the intensity of frustration (Q3) they are able to induce, indicating these responses were distinctly different from one other.


\subsection{Facial Feature Data}

Analysis of participants' facial feature data showed that their expressions during baseline, correct and recovery system behaviors, and unexpected system behaviors were statistically different from each other. This implies that computer vision can be used to detect human responses to subtle non-critical events (e.g., stimuli).
Specifically, the significant differences between \textit{surprise} and \textit{confusion} was seen from certain AUs, such as AU45 (Blink), AU07 (Lid Tightener), and AU10 (Upper Lip Raiser). Other significant features are around the mouth region, such as AU14 (Dimpler), AU25 (Lips Part), AU12 (Lip Corner Puller), AU23 (Lip Tightener), and AU17 (Chin Raiser). We recommend that readers refer to the supplementary video to observe these distinct facial expressions in context and better understand the nuanced differences between these emotional responses.

These results are consistent with prior work, as \cite{ihme2018frustration} pointed out some important features for emotion detection that are near the mouth region, such as AU10, AU12, AU17, and AU23. \cite{zepf2019towards} pointed out that AU04, AU10, and AU14 have previously been associated with negative emotions and frustration. The inclusion of smile-related features such as AU06, AU12, and AU26 are signs of frustration, as people might start smiling when they are frustrated \cite{hoque2011acted, hoque2012exploring}. Moreover, AU45 (Blink) was one of the most significant characteristics to differentiate between triggered responses. This is consistent with existing works, as blink could potentially link different levels of frustration \cite{crnovrsanin2014stimulating}. These results imply that participants' facial expressions during subtle non-critical events have similar characteristics to existing works in other contexts.
Note that there were people who only reacted slightly or even did not react at all, along with people who had stronger response signals. The above results are aggregated across all participants. Although the results of the stimulus check revealed only small differences between confusion and frustration, we observed significant differences between them from facial features.

The distinct differences in facial expressions for the three responses suggest that different types of unexpectedness might also elicit distinct patterns of facial responses. For instance, a repeated system error (which could lead to frustration) might trigger a different combination and intensity of AUs compared to an unexpected piece of information presented by the system (potentially leading to surprise or confusion). This differentiation between surprise, confusion, and frustration based on facial cues implies that the underlying cognitive and emotional appraisals of different unexpected events manifest in observable facial expressions. This opens the possibility of using facial expression analysis to not only identify the user's affective state but also to potentially infer the nature of the unexpected event they are experiencing in real-time. For example, when detecting expressions of confusion, the system could automatically provide additional explanations about its actions, while frustration might trigger a different remediation strategy, such as offering alternative options, adapting its behavior, or initiating interactive system learning from human feedback. 

However, our study has its limitations.  First, our study was conducted in a controlled laboratory setting, which may not fully capture the complexity of real-world driving scenarios and lighting conditions. Second, individual differences in facial expressivity and cultural differences may impact the generalizability of our findings. Finally, the specific stimuli used in our study represent only a subset of possible unexpected behaviors that could occur in autonomous vehicles. 

%% file: text/8_conclusion.tex
\section{Conclusion}


We conducted a human study with a driving simulator to collect a multimodal dataset of subtle emotional responses to noncritical unexpected in-vehicle behaviors. Our crafted scenarios successfully triggered surprise, confusion, and frustration in the participants, which were confirmed through self-reported survey ratings on the perception of the system's behavior and the participants' facial expressions. Future work will involve a deeper exploration of other camera positions to determine if human behavior can be accurately identified using other camera views. 
We believe that new algorithms will be needed to enable viewpoints partially obstructed by lower head angles. We will investigate if the verbal data collected can enhance performance. We hope to leverage verbal and visual cues to develop a real-time detector of unexpected vehicle behavior and explore adaptive vehicle strategies that could be designed to improve user experience and trust in autonomous vehicles. 